\documentclass[a4paper,11pt]{article}
\usepackage{pos}
\usepackage{array, booktabs}
\usepackage[demo, export]{adjustbox}  % it also load graphicx
\usepackage{subcaption}
\usepackage[skip=1ex]{caption}

\title{Neutralina: promoting science and gender-equality in Latin-America}
\author*[a]{Lucia Ximena Coll Saravia}

\affiliation[a]{Deutsches Elektronen-Synchrotron DESY, Notkestr. 85, 22607 Hamburg, Germany}

\emailAdd{lucia.ximena.coll.saravia@cern.ch}

\abstract{The Covid-19 pandemic has exposed certain societal weaknesses, including the lack of scientists in the media and the readiness of the public to believe in fake news. "Neutralina" is a science communication personality created on Instagram (@neutralina.lu) in response to the observed need for scientific outreach done by women in Peruvian and Latin American society. The objectives of this project include normalizing the presence of women in science, fighting against stereotypes and fake news, and disseminating scientific knowledge. Neutralina has attracted a sizable young audience and is expanding beyond Instagram into other formats such as podcasts, real-life conferences, and roundtable discussions. Statistics detailing its growth will be presented, alongside strategies employed to engage the young audience.}

\FullConference{%
  42nd International Conference on High Energy physics - ICHEP2024\\
  17-24 July, 2024\\
  Prague, Czech Republic
}

\begin{document}
\maketitle

\section{Motivation}
This project is inspired by the experiences of growing up and studying in Lima, Peru, where science was not commonly seen as a career path, especially for women. Throughout education, both peers and teachers showed little interest in science, and career counseling rarely included it as an option. Many young women, including the project's founder, were unaware that becoming a scientist was a viable possibility. Despite pursuing a degree in physics, with a focus on high-energy physics, the significant gender gap was apparent, with only one female classmate and no female professors.

In 2021, Neutralina was launched on Instagram \cite{myig} during the COVID-19 pandemic, when science and scientists played a crucial role in combating misinformation. Neutralina aims to provide credible scientific information and serve as a role model to help bridge the gender gap in science. The name Neutralina, inspired by the theoretical particle 'neutralino' from the Supersymmetric model, was chosen to resonate with young girls and emphasize that science is for everyone.

\section{Objectives and performance}
This project has three main objectives: promote visibility of women in science and challenge stereotypes, combat misinformation while encouraging critical thinking, and foster curiosity, especially among girls. The aim is to demonstrate that scientists are ordinary people with everyday activities, who can be relatable. Additionally, it seeks to show that individuals from Latin America, with a Latin American education (bachelor's and/or master's), can contribute to advanced scientific research at the forefront of knowledge, both within their own countries and at leading research centers worldwide. 

Launched in 2021, the account now has around 27,000 followers and reaches over 125,000 accounts, with more than 16,000 actively engaging with the content. Over half of the recommended content is actively chosen to be viewed, though it’s unclear how long it is watched. With over 270 posts, mainly consisting of short videos (about 1 minute) and pictures posted at least once a week, the account’s growth is largely driven by a few viral posts, which seem to benefit from Instagram’s unpredictable algorithm.

\begin{figure}[h]
\raisebox{0.2cm}{\includegraphics[width=0.28\textwidth]{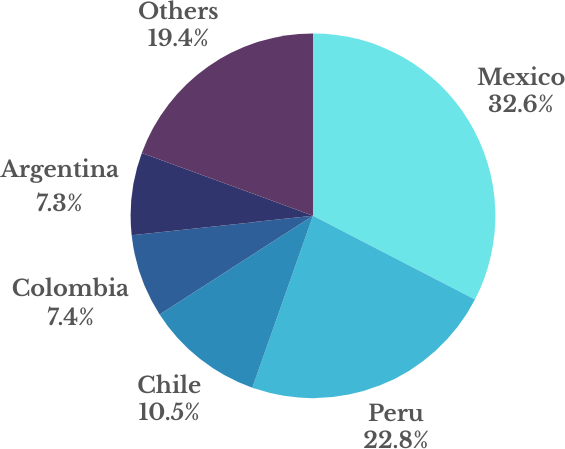}}\hspace{0.2cm}
\raisebox{0.5cm}{\includegraphics[width=0.2\textwidth]{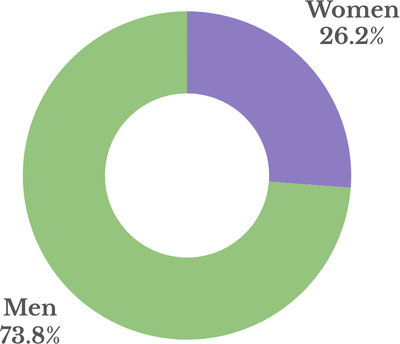}}
\centering
\caption[Demography in the community.]{Demographic statistics present in the community including the countries where the accounts are located and user's gender.  Data collected on June 2024.}
\label{fig:community}
\end{figure}

As of June 2024, the majority of Neutralina’s audience comes from Mexico (33\%) and Peru (23\%), followed by Chile (11\%), Argentina (7\%), Colombia (7\%), and other regions (19\%). Most viewers are between 18 and 35 years old, with about 26\% identifying as women and the rest as men (Figure \ref{fig:community}).

\section{Impact and conclusions}
This project began as an Instagram account and has since expanded beyond social media. Neutralina has engaged in various in-person events, including PUCP Science Talks \cite{pucp-scitalks}, round tables in bars, and interviews \cite{andina}, bringing the online persona to life and deepening connections with the audience. The recent expansion to YouTube \cite{myyt} aims to provide in-depth science content through longer, high-quality videos, reaching a broader and more diverse audience. This expansion, along with in-person events, enhances the project's impact by offering richer content and fostering deeper engagement

\begin{figure}[h]
\begin{subfigure}{0.45\textwidth}
\includegraphics[width=0.8\textwidth]{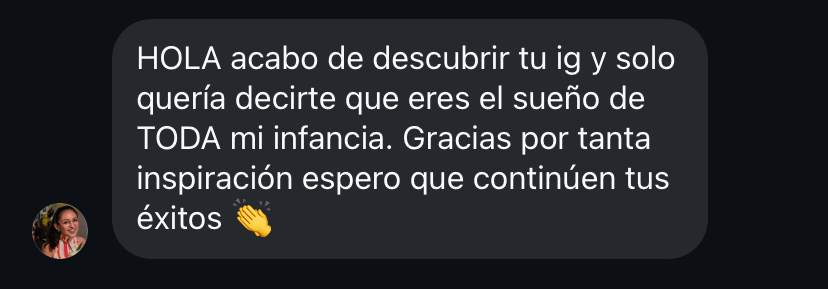} 
\centering
\vspace{0.5em} % Adjusts space between image and caption
\caption{''...I have just discovered your account and just wanted to say that you are my dream of childhood. Thanks for all the inspiration...''}
\label{fig:com2}
\end{subfigure}
\begin{subfigure}{0.45\textwidth}
\includegraphics[width=0.8\textwidth]{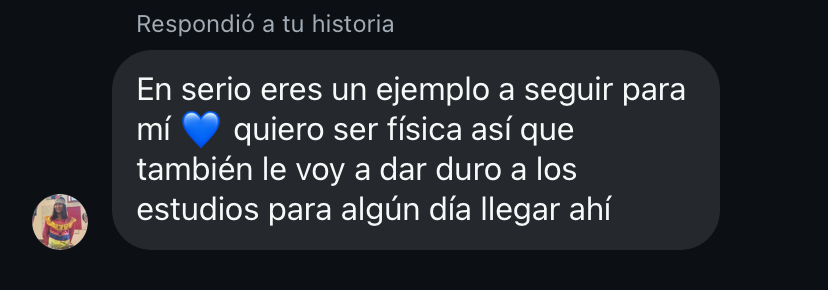} 
\centering
\vspace{0.5em} % Adjusts space between image and caption
\caption{''Seriously, you are a role model for me. I also want to be a physicist, so I will study hard to get there someday.''}
\label{fig:com4}
\end{subfigure}
\centering
%\vspace{1em} % Adjusts space between the two subfigures' captions
\caption[Comments]{Direct messages received in the Instagram account. English translation shown in captions.}
\label{fig:comments}
\end{figure}

While the project's success is reflected in growing follower numbers, social media reach, and engagement metrics, its true impact lies in fostering a meaningful connection with its audience. Neutralina has inspired young women across Latin America, particularly in Peru, to view science as an exciting and attainable career path, as illustrated by direct messages shown in Figure \ref{fig:comments}.

Led by a Peruvian physicist pursuing a PhD at DESY, the project represents a commitment to improving opportunities for girls in Latin America to pursue scientific careers. By humanizing the scientist's role and demonstrating that advanced research is feasible both within and beyond the region, Neutralina continues to challenge stereotypes, combat misinformation, and inspire curiosity in the next generation of scientists

\end{document}